\documentclass[doublecol,figures]{epl2}
\usepackage{graphicx,amsmath}
\def\be{\begin{eqnarray}}
\def\ee{\end{eqnarray}}
\def\bc{\begin{center}}
\def\ec{\end{center}}

\def\rmd{{\rm d}}

\newcommand{\lsim}{\stackrel{\scriptstyle <}{\phantom{}_{\sim}}}

\title{Running condensate in moving superfluid}
\author{E.E. Kolomeitsev\inst{1} \and  D.N. Voskresensky\inst{2}}
\shortauthor{E.E. Kolomeitsev and D.N. Voskresensky}
\institute{\inst{1}Matej Bel  University, SK-97401 Banska Bystrica, Slovakia\\
\inst{2}National Research Nuclear University (MEPhI), 115409 Moscow, Russia}
\abstract{A possibility of the condensation of excitations with non-zero momentum
in moving superfluid media is considered in terms of the Ginzburg-Landau model.
The results might be applicable to the superfluid $^4$He, ultracold atomic Bose gases,
various superconducting and neutral systems with pairing, like ultracold atomic Fermi
gases and the neutron component in compact stars. The order parameters, the energy gain,
and critical velocities are found.
}
\date{\today}
\begin{document}

\maketitle

\section{Introduction}

A possibility of the condensation of rotons in the He-II, moving in
a capillary at zero temperature with a flow velocity exceeding the
Landau critical velocity, was suggested in~\cite{Pitaev84}.
In~\cite{Voskresensky:1993uw} the condensation of excitations with
non-zero momentum in various moving non-relativistic and
relativistic media (not necessarily superfluid) was studied. A
possibility of the formation of a condensate of zero-sound-like
modes with a finite momentum in normal Fermi liquids at non-zero
temperature was further discussed in~\cite{Vexp95}. The
condensation of excitations in cold atomic Bose gases was recently
studied in~\cite{BP12}.  The
works~\cite{Pitaev84,Voskresensky:1993uw,BP12} disregarded a
non-linear interaction between the ``mother'' condensate of the
superfluid and the condensate of excitations. The condensation of
excitations in superfluid systems at finite temperatures, {\it
i.e.}, in the presence of a normal subsystem, also has not  been
studied yet.

The key idea is as follows~\cite{Pitaev84,Voskresensky:1993uw}.
When a medium moves as a whole with respect to a laboratory frame
with a velocity higher than a certain critical velocity, it may
become energetically favorable to transfer a part of its momentum
from particles of the moving medium to a condensate of Bose
excitations with a non-zero momentum $k\neq 0$. It would happen,
if the spectrum of excitations is soft in some region of the
momenta. Whether the system is moving linearly with a constant
velocity or it is resting, is indistinguishable according to the
Galilei invariance. Thus, there should still exist a physical
mechanism allowing to produce excitations. The excitations can be
created near a wall, which singles out the laboratory reference
frame, or they can be produced by interactions among particles of
the normal subsystem at non-zero temperature. They can be also
generated provided the motion is non-inertial, {\it e.g.}, in the
case of a rotating system.

In the He-II there exists a branch of roton
excitations~\cite{LP1981,GS}. The typical value of the energy of
the excitations $\Delta_{\rm r}=\epsilon (k_{\rm r})$ at the roton
minimum for $k=k_{\rm r}$ depends on the pressure and temperature.
According to~\cite{BrooksDonelly}, for the saturated vapor pressure
$\Delta_{\rm r}= 8.71$\,K at $T= 0.1$\,K and
$\Delta_{\rm r}= 7.63$K  at $T=2.10$\,K, and $k_{\rm r} \simeq 1.9\cdot 10^{8}\hbar/{\rm
cm}$ in the whole temperature interval. (We
put the Boltzmann constant $k_{\rm B}=1$). An appropriate branch
of excitations may exist also in normal Fermi
liquids~\cite{Vexp95} and in cold Bose~\cite{BP12,coldatmode} and
Fermi~\cite{Kagan} atomic gases. In neutral Fermi liquids with the
singlet pairing, characterized by the pairing gap $\Delta$, there
exist~\cite{Kulik43} the low-lying Anderson-Bogoliubov mode of
excitations with $\epsilon (k)=kv_{\rm F}/\sqrt{3}$ for $k\to 0$
and $\epsilon(k)\to 2\Delta$ for large $k$ ($k\lsim 2p_{\rm F}$),
and the Schmid mode $\epsilon (k)\simeq 2\Delta$. Here $p_{\rm F}$
is the Fermi momentum, $v_{\rm F}$ is the Fermi velocity. In
charged Fermi liquids with the singlet pairing there is also the
suitable low-lying Carlson-Goldman mode starting at zero energy
for a small momentum and reaching the value $\epsilon = 2\Delta$
for large $k$.

Below, we study a possibility of the condensation of excitations
in the state with a non-zero momentum in moving media in the presence
of the superfluid subsystem. The systems of our interest are
neutral bosonic superfluids, such as the superfluid $^4$He and
cold Bose atomic gases, and systems with the Cooper pairing, like
the neutron liquid in neutron star interiors, cold Fermi atomic
gases or charged superfluids, as paired protons in neutron star
interiors and paired electrons in metallic superconductors. In
contrast to previous works we take into account that the
superfluid subsystem and the bosonic excitations should be
described in terms of the very same macroscopic wave-function.
Also, our consideration is performed for non-zero temperature,
{\it i.e.}, the presence of the normal component is taken into account.

\section{Ginzburg-Landau functional}

We start with expression for the Ginzburg-Landau (GL) free-energy
density of the superfluid subsystem in its rest reference frame
for the temperature $T<T_c$, \cite{LP1981,GS}:
\begin{align}
&F_{\rm GL}[\psi]=\frac{c_0}{2}|\hbar\nabla\psi|^2-a(t)\,|\psi|^2
+\frac{1}{2} b(t) |\psi|^4\,, \nonumber\\
&a(t)=a_0\,t^\alpha\,,\,\,  b(t)=b_0\,t^\beta\,,\,\, t=(T_c -T)/T_c\,.
\label{FGL}
\end{align}
Here $T_c$ is the critical temperature of the second-order phase
transition, and $c_0$, $a_0$ and $b_0$ are phenomenological
parameters. When treated within the mean-field approximation,
the functional $F_{\rm GL}$ should be an analytic function of $t$.
Then, from the Taylor expansion of $F_{\rm GL}$ in $t$ it follows that $\alpha =1$,
$\beta =0$. The width of the so-called fluctuation region, wherein
the mean-field approximation is not applicable, is evaluated
from the Ginzburg criterion: in this region of temperatures in the
vicinity of $T_c$, long-range fluctuations of the order parameter
are mostly probable, i.e. their probability is
$W\sim e^{-F^{\rm eq}_{\rm GL}V_{\rm fl}/T}\sim 1$, where $F^{\rm
eq}_{\rm GL}$ is the equilibrium value, $V_{\rm fl}\sim \xi^3$ is
the minimal volume of the fluctuation of the order parameter,  the
coherence length $\xi$ is the typical length scale characterizing
the order parameter.

For metallic superconductors the fluctuation region proves to be very narrow and
the mean-field approximation holds for almost any temperatures below $T_c$,
except a tiny vicinity of $T_c$. Thus, neglecting this narrow
fluctuation region one may use $\alpha =1$, $\beta =0$ in
(\ref{FGL}). For the fermionic systems with the singlet pairing, in
the weak-coupling (BCS) approximation the parameters can be extracted from the
microscopic theory~\cite{LP1981}:
\begin{align}
c_0={1}/{2m^*_{\rm F}}\,, \,\,
a_0 ={6\pi^2 T_c^2}/{(7\zeta(3)\mu)}\,, \,\,
b_0={a_0}/{n}\,,
\label{BCS-param}
\end{align}
where $m^*_{\rm F}$ stands for the effective fermion mass
($m^*_{\rm F}\simeq m_{\rm F}$ in the weak-coupling limit),
$n=p_{\rm F}^3/(3\pi^2\hbar^3)$ is the particle number density,
and the fermion chemical potential is $\mu\simeq \epsilon_{\rm F}=
{p_{\rm F}^2}/{(2m^{*}_{\rm F})}$. The function $\zeta(x)$ is the
Riemann $\zeta$-function and $\zeta (3)=1.202$. The values of the
parameters are obtained for $t\ll 1$. With these BCS parameters we
have $|\psi |^2=nt$ and the pairing gap $\Delta =
T_c\sqrt{\frac{8\pi^2 t }{7\zeta (3)}}$, see~\cite{Abrikosov}.

For He-II the fluctuations prove to be important for all
temperatures below $T_c$~\cite{GS}. Including long-range
fluctuations, the coefficients of the Ginzburg-Landau functional
are now renormalized:  due to a divergency  (logarithmic or
power-law-like) of the specific heat at the critical temperature
$T_c$, quantities $a$ and $b$ in eq.~(\ref{FGL}) become
non-analytic functions of $t$ with non-integer $\alpha$ and
$\beta$. When the contributions of long-range fluctuations are
completely taken into account the Ginzburg parameter $F^{\rm
eq}_{\rm GL}\xi^3$ must become independent of the temperature.
Since $F^{\rm eq}_{\rm GL}\propto t^{2\alpha -\beta}$ and $\xi
\propto t^{-\alpha/2}$, we obtain $\alpha/2-\beta =0$. Using the
experimental fact that the specific heat of the He-II contains no
power divergence at $T\to T_c$, we get $2\alpha -\beta -2 =0$.
From these two relations we find $\alpha =4/3$, $\beta =2/3$.
Other parameters of He-II at the saturated vapor pressure are
\cite{GS}: $T_c =2.17$\,K, $a_0/ T_c^{4/3} =1.11\cdot 10^{-16}{\rm
erg/K^{4/3}}$, $b_0/ T_c^{2/3} =3.54\cdot 10^{-39}{\rm erg\cdot
cm^3/K^{2/3}}$. This parameterization holds for $10^{-6}<t<0.1$,
but for rough estimates can be used up to $t=1$. {\it E.g.}, using
eq.~(\ref{FGL}), with the helium atom mass  $m=6.6\cdot
10^{-24}$\,g we evaluate  the He-II mass-density as $m
a_0/b_0\simeq 0.3\,{\rm g/cm^{3}}$, which is of the order of the
experimental value  $\rho_{\rm He}=0.15\,{\rm g/cm^{3}}$ at $P=
0$.

\section{Moving cold superfluid}

We consider now a superfluid moving with a constant velocity
$\vec{v}$ parallel to a wall. The latter singles out the
laboratory frame and an interaction of the fluid with the wall may
lead to creation of excitations in the fluid. We start with the
case of vanishing temperatures. The whole medium is superfluid and the
amplitude of the order parameter can be related to the particle
density $\rho_s=m\,n = m\,|\psi_{\rm in}|^2=m\, a_0/b_0$,
$\psi_{\rm in}$ is the order parameter in the absence of the
excitations (``in''-state). The energy of the medium in the
laboratory frame is $E_{\rm in}=m n v^2/2 - b_0 n^2/2$\,.

When the speed of the flow, $v$, exceeds the Landau critical
velocity, $v_c^{\rm L}$, near the wall there may appear
excitations with the momentum $k=k_0$ and the energy
$\epsilon(k_0)$ as calculated in the rest frame of the superfluid,
where the momentum $k_0$ corresponds to the minimum of the ratio
$\epsilon (k)/k$. For instance, for He-II the spectrum
$\epsilon(k)$ is the standard phonon+roton spectrum, normalized as
$\epsilon(k)\propto k$ for small $k$. The appearance of a large
number of excitations motivates us to assume that for $v>v_c^{\rm
L}$ in addition to the mother condensate the excitations may form
a new condensate with the momentum $k=k_0\neq 0$, (``fin''-state).
The momentum $k_0$ should be found now from minimization of the
free energy. As we have noticed before, in previous
works~\cite{Pitaev84,Voskresensky:1993uw,BP12} it was assumed that
the condensate of excitations decouples from the mother
condensate. Now we are going to take into account that the
condensate of excitations is described by the very same
macroscopic wave-function as the mother condensate. Then the
resulting order parameter $\psi_{\rm fin}$  is the sum of the
contributions of the mother condensate, $\psi$, and  the
condensate of excitations, $\psi'$, i.e., $\psi_{\rm fin} =\psi
+\psi'\,.$ We choose the simplest form of the order parameter for
the condensate of  excitations,
\begin{align}
\label{ex} \psi'=\psi'_0 e^{-i(\epsilon (k_0)t - \vec{k}_0\vec{r})/\hbar},
\end{align}
with the amplitude $\psi'_0$, being constant for the case of the
homogeneous system that we consider. The particle number conservation yields
\begin{align}
\label{n} &n=|\psi+\psi'|^2=|\psi|^2+|\psi'_0|^2+\dots\,.
\end{align}
Here ellipses stand for the spatially oscillating term, which
vanishes after the averaging over the system volume.

As the particle density, the initial momentum density is
redistributed in our case between the fluid and the condensate of
excitations:
\begin{align}
\label{momentumcons-n}
\rho_s\, \vec{v}=(\rho_s - m\,|\psi'_0|^2)\, \vec{v}_{\rm fin}
+(\vec{k}_0+m\vec{v}_{\rm fin})\, |\psi'_0|^2\,.
\end{align}
The energy of the moving matter in the presence of the condensate
of excitations is
\begin{align}
E_{\rm fin}=\frac{\rho_s v_{\rm
fin}^2}{2}+(\epsilon(k_0)+\epsilon_{\rm
bind})\,|\psi'_0|^2-a_0\,|\psi|^2+\frac{b_0 n^2}{2}\,,
\label{Efin-1}
\end{align}
where the energy of the excitation $\epsilon(k)$ should be
counted from the binding energy of a particle in the condensate at
rest $\epsilon_{\rm bind}=\partial E_{\rm in}(v=0)/\partial
n=-b_0n=-a_0$\,. Replacing
eq.~(\ref{momentumcons-n}) in~(\ref{Efin-1}) and using
eq.~(\ref{n}) we express the change of the volume-averaged energy
density owing to the appearance of the condensate of excitations,
$\delta \bar{E}=\bar{E}_{\rm fin}-\bar{E}_{\rm in}$, as
\begin{align}
\delta \bar{E}= (\epsilon(k_0) - k_0\,v)|\psi'_0|^2 + {k_0^2|\psi'_0|^4}/{(2\rho_s)}\,.
\end{align}
Minimizing this functional with respect to $\psi'_0$ we obtain the
condensate amplitude
\begin{align}
|\psi'_0|^2=({\rho_s}/{k_0})\,(v-v_c^{\rm L})\,\theta(v-v_c^{\rm
L}) \,,\,\, v_c^{\rm L} = {\epsilon(k_0)}/{k_0}\,. \label{T0-sol}
\end{align}
From~(\ref{momentumcons-n}) we find that because of condensation
of excitations with $k\neq 0$ the flow is decelerated to the
velocity $v_{\rm fin}=v_c^{\rm L}$.  The volume-averaged energy
density gain due to the appearance of the condensate of
excitations is
\begin{align}\label{deltaE}
\delta \bar{E}= -{\textstyle \frac12}\rho_s(v-v_c^{\rm L})^2\theta(v-v_c^{\rm L})\,.
\end{align}
Minimization of $\delta \bar{E}$ with respect to $k_0$ gives the
condition $\rmd v_c^{\rm L}/\rmd k_0=0$, which is exactly the
condition for the standard Landau critical velocity.  The
condensate of excitations appears by a second-order phase
transition. The amplitude of the  condensate of excitations~(\ref{T0-sol})
grows with the velocity, whereas the amplitude of
the mother condensate decreases. The value $|\psi|^2 $ vanishes
when $v=v_{c2}$, the second critical velocity, at which
$|\psi'_0|^2=n$ according to eq.~(\ref{n}). The value of the
second critical velocity $v_{c2}$ is evaluated from~(\ref{T0-sol})
as $v_{c2}=v_c^{\rm L}+k_0/m$. When the mother condensate
disappears, at $v=v_{c2}$, the excitation spectrum is cardinally
reconstructed, and the superfluidity destruction occurs as a
first-order phase transition. We assume that for
$v>v_{c2}$ the excitation spectrum has no low-lying local minimum
at finite momentum. Then
the amplitude $|\psi'_0|^2$ jumps from  $n$ to 0 and $\delta
\bar{E}$ jumps from $\delta \bar{E}(v_{c2})=-\rho k_0^2/(2m^2)$ to
0 at $v=v_{c2}$. The case, when in the absence of the mother condensate
the spectrum of Bose excitations has a low-lying local
minimum at $k\neq 0$, has been considered in~\cite{Voskresensky:1993uw,Vexp95}.
Note that in practice the
reconstruction of the spectrum  may occur for a smaller velocity
than that we have estimated. For example, for fermionic superfluids it
always should be $|\psi'_0|^2\ll n$, otherwise the Fermi sea
itself would be destroyed.

In moving superfluids there exist excitations of the type of
vortex rings. The energy of the vortex is estimated as
$\epsilon^{\rm vort}=2\pi^2 \hbar^2|\psi|^2 R m^{-1} \ln (R/\xi)$,
see~\cite{GS,Khalatnikov}, and the momentum is $p^{\rm
vort}=2\pi^2\hbar|\psi|^2 R^2$, $m$ here is the mass of the pair
for systems with pairing, and the mass of the boson quasiparticle
in bosonic superfluids, {\it e.g.}, the  mass of the $^4$He atom
in case of the He-II, $R$ is the radius of the vortex ring, and
$\xi\sim c_0^{1/2} a^{-1/2}_0 t^{-\alpha/2}$ is the minimal length
scale associated with the mother condensate. Thus,
$
v_{c1}=\epsilon^{\rm vort}/p^{\rm vort}= \hbar (Rm)^{-1}\ln
(R/\xi)\,,
$ 
is the Landau critical velocity for the vortex production, where
now $R$ is the distance of the order of the  size of the system.
For $v>v_{c1}$ the vortex rings are pushed out of the medium, if
the density profile has even slight inhomogeneity. Note that for
spatially extended systems the value $v_{c1}$ is usually lower
than the Landau critical velocity $v_c^{\rm L}$. The flow moving
with the velocity $v$ for $v_{c1}\leq v \leq v_{c2}$ may be
considered as metastable, since the vortex creation probability is
hindered by a large potential barrier and formation of a vortex
takes a long time~\cite{vortex-creation}. Note that already for
$v$ just slightly exceeding $v_{c1}$, the number of the produced
vortices may become sufficiently large and their interaction
forces the normal and superfluid components to move as a solid,
with the same velocity, even if initially they have had different
velocities. In the exterior regions of  the vortex cores the
superfluidity still persists and our consideration of the
condensation of excitations in the velocity interval $v_{c}^{\rm
L}<v<v_{c2}$ is applicable.

In the presence of the condensate of excitations the density becomes
spatially oscillating around  its averaged value.  For a weak condensate,
{\it i.e.}, $|v-v_c^{\rm L}|\ll v_c^{\rm L}$, we find perturbatively $\delta
n=n-\bar{n}\approx
2\sqrt{n}|\psi'_0|\cos((\epsilon(k_0)\,t-\vec{k}_0\vec{r})/\hbar\,)$. Such
a density modulation predicted in~\cite{Pitaev84} was reproduced
in the numerical simulation of the supercritical flow in He-II using the
realistic density functional~\cite{Ancilotto05}.

The above consideration holds for any superfluid with the
conserved number of particles in the mother condensate plus the
condensate of excitations, {\it e.g.}, for the cold Bose gases, if
the spectrum of the over-condensate excitations is such that
$\epsilon (k_0)/k_0$ has a minimum at $k=k_0\neq 0$, as has been
conjectured in~\cite{BP12}.


In case of the Fermi systems with pairing, for the bosonic modes
(Anderson-Bogoliubov, Schmid and Carlson-Goldman ones) with the
excitation energy $\simeq 2\Delta$, cf.~\cite{Kulik43}, the
maximum momentum, up to which the mode is not yet destroyed, is
$k_0 \simeq 2p_{\rm F}$, cf.~\cite{Kagan,Urban}. Hence, for these
modes
 the Landau critical velocity is $v_c^{\rm L} \simeq
\Delta/p_{\rm F}$, and for $v> v_c^{\rm L}$ there is a chance for
the condensation of the Bose excitations as we considered above.
Besides bosonic excitations there exist fermionic ones with the
spectrum $\epsilon_{\rm f}(p)=\sqrt{\Delta^2+v_{\rm F}^2(p-p_{\rm
F})^2}$\,. Stemming from the breakup of Cooper pairs, the
fermionic excitations are produced pairwise. Therefore, the
corresponding (fermion) Landau critical velocity is $v_{c,{\rm
f}}^{\rm L}=\min_{\vec{p}_1,\vec{p}_2} [(\epsilon_{\rm
f}(p_1)+\epsilon_{\rm f}(p_2))/|\vec{p}_1+\vec{p}_2|]$\,. The
latter expression reduces to~\cite{Castin14} $v_{c,{\rm f}}^{\rm
L}=(\Delta/p_{\rm F})/(1+\Delta^2/p_{\rm F}^2 v_{\rm
F}^2)^{1/2}$\,. We see that up to a small correction  $v_{c,{\rm
f}}^{\rm L}\simeq v_c^{\rm L}$. For $T\to 0$ and  $0\le v/v_c^{\rm
L}-1\ll1$ we find $<\vec{p}\,\vec{v}>/\rho \simeq 2\sqrt{2v_c^{\rm
L}}(v-v_c^{\rm L})^{3/2}$ and  the energy gain due to the fermion
pair breaking is
\begin{align}\label{deltaEf}
\delta \bar{E}_{\rm
pair}&=\intop\frac{2\rmd^3p}{(2\pi\hbar)^3}({\epsilon_{\rm
f}(p)-\vec{p}\,\vec{v}}\,)\theta (\epsilon_{\rm
f}(p)-\vec{p}\,\vec{v}\,)\nonumber\\
 &\approx - 2\sqrt{2}\rho (v- v_{c {\rm
}}^{\rm L})^{2} [v/v_{c {\rm }}^{\rm L}-1]^{1/2}\,.
\end{align}
Moreover, for $v>v_{c,{\rm f}}^{\rm L}$ the pairing gap decreases
with increase of $v$ as~\cite{Zagozkin} $\Delta(v)/\Delta\approx
1-(3/2)(v/v_c^{\rm L}-1)^2$, reaching zero for $v=v_{c2,{\rm
f}}^{\rm L}=\frac{e}{2}v_c^{\rm L}$ (Rogers-Bardeen
effect~\cite{Bardeen}). The energy gain~(\ref{deltaEf}) is less
than~(\ref{deltaE}) and the production of the condensate of Bose
excitations is energetically more profitable than the Cooper pair
breaking. Since in the presence of the condensate of excitations
$v_{\rm fin}=v_{c}^{\rm L}$ an additional energy gain due to the
appearance of the latter is: ${F}_{\rm GL }^{\rm
eq}(T=0,\Delta)-{F}_{\rm GL }^{\rm eq}(T=0,\Delta (v))\approx
-(9/8)\rho (v-v_c^{\rm L})^2$, for $0\le v/v_c^{\rm L}-1\ll 1$.
For $v>v_{c2,{\rm f}}^{\rm L}$ the gain becomes ${F}_{\rm GL
}^{\rm eq}(T=0,\Delta)=-3\rho ( v_c^{\rm L})^2/4$.

\section{Moving  superfluid-normal composites}

We turn now to the case of systems consisting of normal and
superfluid parts, like He-II at finite temperature, metallic
superconductors, or neutron star matter. Here, the number of
particles in the condensate is not conserved even at $v=0$ since a
part of particles can be transferred from the superfluid to the
normal subsystem. The state of the superfluid subsystem is
described by the GL functional~(\ref{FGL}). We assume that
\emph{the normal and superfluid subsystems move with the same velocity
$\vec{v}$} with respect to a laboratory frame. This means that we
consider the motion of the fluid during the time $\tau$ shorter
than the typical friction time, $\tau_{\rm fr}^{\rm norm}$, at
which the normal component is decelerated, if the fluid has a
contact with the wall.

Minimization over the order parameter in the rest frame of the fluid yields
\begin{eqnarray}
|\psi^{\rm eq}_{v=0}|^2={a(t)}/{b(t)} \,, \,\, F_{\rm GL}^{\rm
eq}[\psi^{\rm eq}_{v=0}]= -{b(t)}|\psi^{\rm eq}_{v=0}|^4/2\,.
\label{old-cond}
\end{eqnarray}

In the absence of the population of excitation modes the superfluid
and normal subsystems decouple. In this case the initial
free-energy density of the system is given by
\begin{align}
F_{\rm in} = {\rho v^2}/{2}+F_{\rm bind} - {a^2(t)}/({2\,
b(t)})\,.
\end{align}
Here $\rho$ is the total (normal+superfluid) mass density,
$\rho =\rho_n (v=0)+m\,|\psi (v=0)|^2$, and $F_{\rm bind}$ is a
binding free-energy density of the normal subsystem in its rest
reference frame (coinciding with the rest frame of the superfluid
in the case under consideration). The explicit form of $F_{\rm
bind}$ is not of our interest.

When the condensate of excitations is formed, the initial momentum density
is redistributed between the fluid and the condensate of excitations:
\begin{align}
\label{momentumcons}
\rho\, \vec{v}=(\rho- m\,|\psi'_0|^2)\, \vec{v}_{\rm fin}
+(\vec{k}_0+m\vec{v}_{\rm fin})\, |\psi'_0|^2\,.
\end{align}
Here  we assume (as argued below) that after the appearance of the
condensate of excitations in the form (\ref{ex}) the normal and
superfluid subsystems continue to move with one and the same
velocity $\vec{v}_{\rm fin}$.  In the presence of the condensate
of excitations the free energy density becomes
\begin{align}
F_{\rm fin}=&{\textstyle \frac12}\rho\,v_{\rm fin}^2 + F_{\rm bind} +
F_{\rm GL}[\psi,\nabla \psi =0]\\ +&(\epsilon
(k_0)+\epsilon_{\rm bind})\,|\psi'_0|^2 + 2\,b(t)\, |\psi|^2\,|\psi'_0|^2 + {\textstyle \frac12}
b(t) |\psi'_0|^4\,.\nonumber
\end{align}
To get this expression we used eq.~(\ref{FGL}) and replaced there
$\psi$ with $\psi_{\rm fin}=\psi+\psi{'}$. The energy of
excitations should be counted here from the excitation energy on
top of the mother condensate at rest determined by
eq.~(\ref{old-cond}),
$\epsilon_{\rm bind} = {\partial}F_{\rm
GL}^{\rm eq} [\psi=\psi_{v=0}^{\rm eq}+\psi']/{\partial|\psi'|^2}\big|_{\psi'=0}
=-2a(t)
$\,.

Now, using the momentum conservation (\ref{momentumcons}) we
express $\vec{v}_{\rm fin}$ through $\vec{v}$ and get for the
change of the volume-averaged free-energy density associated with the
appearance of the condensate of excitations,
\begin{align}
\delta \bar{F}& ={\textstyle \frac12}b(t)\big(|\psi|^2-{a(t)}/{b(t)}\big)^2
+\big(\epsilon (k_0) - k_0\,v\big)|\psi'_0|^2 \nonumber\\
&+2\,b(t)\,
\big(|\psi|^2-{a(t)}/{b(t)}\big)\,|\psi'_0|^2 +
{\textstyle \frac12} \widetilde{b}(t)
|\psi'_0|^4\,, \label{Fgen}
\end{align}
where  $\widetilde{b}(t)=b(t) + k_0^2/\rho$ and we put
$\vec{k}_0\parallel \vec{v}$. We note that the normal subsystem
serves as a reservoir of particles  at the formation of the
condensates, which amplitudes are chosen by minimization of the
free energy of the system. Therefore, we vary $\delta \bar{F}$
with respect to $\psi$ and $\psi'_0$ independently. Thus,
minimizing~(\ref{Fgen}) we find
\begin{align}
&|\psi'_0|^2 =\frac{k_0 \big(v-v_c^{\rm L}\big)}{k^2_0/\rho-3b(t)}
\theta\big(v-v_c^{\rm L}\big)\,\theta\big(k_0^2/\rho-3b(t)\big) \,
\,,  \label{orderp}\\
&|\psi|^2 = \big({a(t)}/{b(t)}  - 2|\psi'_0|^2\big)\, \theta (\widetilde{T}_c(v) -T)\theta
(v_{c2}(t)-v)\,.
\nonumber
\end{align}
The quantity $\widetilde{T}_c$ stands for the renormalized
critical temperature, which depends now on the flow velocity, and
$v_{c2}(t)$ stands for the second critical velocity depending on
$T$. The condition $|\psi|^2=0$ implies the relation between $v$ and $T$
\begin{align}\label{tildet}
v=v_c^{\rm L} + {a(t) k_0}/{(2b(t)\rho)}-{3a(t)}/{(2k_0)}\,.
\end{align}
The solution of this equation for the velocity, $v_{c2}(t)$, increases with the
decreasing temperature, and the solution for the temperature, $\widetilde{T}_c (v)$,
decreases with increasing $v$. At $T=\widetilde{T}_c (v)$ or $v=v_{c2}(t)$
we have $|\psi|^2=0$ but $|\psi_0^{'}|^2\neq 0$, and
for $T>\widetilde{T}_c (v)$ or for $v>v_{c2}(t)$ the condensate
$|\psi_0^{'}|^2$ vanishes, if, as we assume, for $|\psi|^2=0$ the
spectrum of excitations does not contain a low-lying branch. Thus,
the superfluidity is destroyed at $T=\widetilde{T}_c (v)$ or
$v=v_{c2}(t)$ in a first-order phase transition.

From (\ref{momentumcons}) and (\ref{orderp}) we find for
$v>v_c^{\rm L}$ and $k^2_0/(\rho b(t))>3$ the resulting velocity
of the flow
\begin{align}\label{vfin}
v_{\rm fin} = v_c^{\rm L} - (v - v_c^{\rm L})/ \big({k_0^2}/{(3 b(t)\rho)}-1\big)
<v_c^{\rm L}\,.
\end{align}

Substituting the order parameters from (\ref{orderp}) in
(\ref{Fgen}), we find for the volume-averaged free-energy density
gain owing to appearance of the condensate of excitations
\begin{align}\label{deltaF}
\delta\bar{F} =- \frac{\rho}{2} \frac{(v-v_c^{\rm
L})^2}{1 - {3b(t)\rho}/{k_0^2}} \theta(v - v_c^{\rm
L})\, \theta(v_{c2}-v)
\end{align}
for $k_0^2/\rho>3b(t)$\,. Thus, for $v_c^{\rm L}<v<v_{c2}$ the
free energy decreases owing to the appearance of the condensate of
excitations with $k\neq 0$ in the presence of the non-vanishing  mother
condensate. The value of $k_0$ is to be found from the
minimization of eq.~(\ref{deltaF}). As $\widetilde{T}_c$, the
condensate momentum $k_0$ gets renormalized and differs now from
the value corresponding to the minimum of $\epsilon (k)/k$. For
$k_0^2/\rho \gg 3{b}_0$ the expression~(\ref{deltaF}) for the gain
in the volume-averaged free-energy density transforms at $T=0$
into the expression~(\ref{deltaE}) for the gain in the
volume-averaged energy density. As in the case of $T=0$, for $T\neq 0$
the condensate of excitations appears at $v=v_c^{\rm L}$ in the
second-order phase transition but it disappears at $v=v_{c2}$ in
the first-order phase transition with the jumps
\begin{align}
\label{jumpvc2} {\delta \bar{F}(v_{c2})}\approx \frac{a^2(t)
k_0^2}{8b^2(t)\rho}, \quad
|\psi'_0(v_{c2})|^2=\frac{a(t)}{2b(t)}\,.
\end{align}

The dynamics of the condensate of excitations is determined by the
equation
 $\Gamma\dot{\psi'} = -{\delta
\bar{F}}/{\delta \psi'^{*}}\,. $
 We emphasize that the above
consideration assumes that the formation rate $\Gamma$ of the
condensate of excitations is faster than the deceleration rate
$1/\tau^{\rm norm}_{\rm fr}$ of the normal subsystem.

When a homogeneous fluid flowing with $v>v_c^{\rm L}$ at
$T>\widetilde{T}_c (v)$ is cooled down to $T<\widetilde{T}_c (v)$,
it consists of three components: the normal and superfluid ones
and the condensate of excitations, all moving with $v_{\rm
fin}<v_c^{\rm L}$. If the system is then re-heated to
$T>\widetilde{T}_c (v)$, the superfluidity and the condensate of
excitations vanish and the remaining normal fluid consists of two
fractions: one is moving with $v_{\rm
fin}(\widetilde{T}_c)<v_c^{\rm L}$ and the other one,
$\delta n=a(\widetilde{T}_c)/(2b(\widetilde{T}_c))$, is moving
with a higher velocity until a new equilibrium is established.
Note also that for fermion superfluids at $T\neq 0$  after the
condensate of excitations is formed the  flow velocity $v_{\rm
fin}< v_{c,{\rm f}}^{\rm L}$, for $v-v_{c}^{\rm L}>4tv_{c}^{\rm
L}/9$, and thereby  the Cooper pair breaking does not occur,
whereas the condensate of Bose excitations is preserved.

\section{Estimates for fermionic superfluids and He-II}

We apply now  expressions derived in the previous section to
several practical cases. First, we consider a fermion system with
pairing. With the BCS parameters~(\ref{BCS-param}) we estimate
$b_0\rho/k_0^2 = 3\Delta^2/(8v_{\rm F}^2p_{\rm F}^2)$ and
$a_0/k_0=3\Delta^2/(4v_{\rm F} p_{\rm F}^2)\,, $ where  $\rho
\simeq nm_{\rm F}$. We see that inequality $k_0^2/\rho \gg 3{b}_0$
is reduced to inequality $\Delta \ll \epsilon_{\rm F}$, which is
well satisfied. In this limit $|\psi_0^{'}|^2$ given by
eq.~(\ref{orderp}) gets the same form as eq.~(\ref{T0-sol}). The
resulting flow velocity after condensation,~(\ref{vfin}), is lower
than the Landau critical velocity but close to it, $v_{\rm fin}
\simeq v_c^{\rm L}-9(v_c^{\rm L})^2(v-v_c^{\rm L})/(8v_{\rm
F}^2)$\,.

Since for the BCS case we have $\alpha =1$, $\beta =0$,  eq.~(\ref{tildet})
for the new critical temperature is easily solved, for $v>v_c^{\rm L}$
\begin{align}
\label{BCST}
\frac{\widetilde{T}_c}{T_c} &= 1 -\frac{2k_0 b_0
(v-v_c^{\rm L})}{a_0(k_0^2/\rho - 3b_0)}
\approx 1-\frac{v-v_c^{\rm L}}{v_{\rm F}}\,.
\end{align}
We also estimate the maximal second critical velocity as
$v_{c2}^{\rm max}\simeq v_c^{\rm L} + v_{\rm F}$.

\begin{figure}
\onefigure[width=5.2cm]{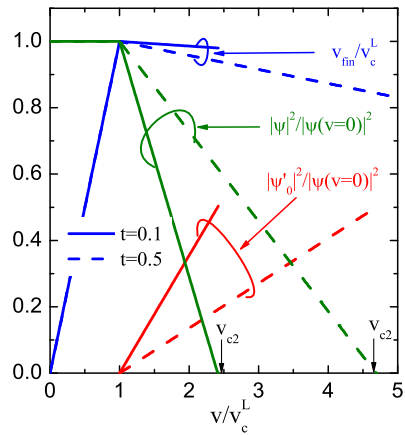}
\caption{Condensate amplitudes
$|\psi|^2$ and $|\psi'_0|^2$, eq.~(\ref{orderp}), and the final
flow velocity $v_{\rm fin}$, eq.~(\ref{vfin}), in superfluid
$^4$He plotted as functions of the flow velocity for various
temperatures, $t=(T_c-T)/T_c$. Vertical arrows indicate the values
of the second critical velocity $v_{c2}$. Velocities are scaled by
the values of the Landau critical velocities $v_c^{\rm
L}(t=0.5)=59\,{\rm m/s}$ and $v_c^{\rm L}(t=0.1)=55\,{\rm m/s}$,
and the condensates are normalized to the condensate amplitude in
the superfluid at rest, eq.~(\ref{old-cond}). } \label{fig:hecond}
\end{figure}

We turn now to the bosonic supefluid -- helium-II. Making use of
the values of the GL parameters presented above and taking into
account that we deal with the rotonic excitation, {\it i.e.},
$k_0\simeq k_{\rm r}$  and $\epsilon(k_0)\simeq \Delta_{\rm r}$,
we estimate, ${k_0^2}/{(b_0\,\rho)}\simeq 47$\,,\, $v_c^{\rm
L}(T\to 0)\simeq 60\,{\rm {m}/{s}}\,,$\, ${a_0}/{k_0}\simeq 16{\rm
{m}/{s}}\,.$ Using the results of~\cite{BrooksDonelly} the
temperature dependence of $v_c^{\rm L}$ can be fitted with 99\%
accuracy as $v_c^{\rm L}(T)/v_c^{\rm L}(0)\simeq
1-0.7e^{-2.14/\tilde{t}}+200\tilde{t} e^{-8/\tilde{t}}$, where
$\tilde{t}=T/T_c$. Using these parameters we evaluate the
condensate amplitudes and the final flow velocity as functions of
temperature and depict them in Fig.~\ref{fig:hecond}. The
condensate of  excitations appears at $v=v_c^{\rm L}$ in a
second-order phase transition. For $v>v_c^{\rm L}$ the amplitude
of the condensate $|\psi'_0|^2$ ($|\psi|^2$) increases (decreases)
linearly with $v$. The closer $T$ is to $T_c$, the steeper  the
change of the condensate amplitudes is. The final velocity of the
flow, which sets in after the appearance of the condensate of
excitations, decreases with the  increase of $v$. For He-II, we
have $\alpha =4/3$, $\beta =2/3$ and the renormalized critical
temperature determined by eq.~(\ref{tildet}) is for $v>v_c^{\rm
L}$:
\begin{align}
\label{TtildeHe}
\frac{\widetilde{T}_c}{T_c} &=
1-\Bigg[\frac{k_0^2}{6b_0\rho}
-
\sqrt{
\frac{k_0^4}{36b_0^2\rho^2} - \frac{2k_0}{3 a_0}\Big[
v-v_c^{\rm L}\Big]}
\Bigg]^{3/2}
\nonumber\\
&\approx
1 - 0.05\,(v/v_c^{\rm L}(T_c)-1)^{3/2}\,.
\end{align}
The mother condensate $|\psi|^2$ vanishes when $v$ reaches the
value of  the second critical velocity $v_{c2}$, which depends on
the temperature as $v_{c2}\approx v_c^{\rm L}(t) + (363
t^{2/3}-23.5t^{4/3}){\rm m/s}$. At $v=v_{c2}$ the superfluidity
disappears in a first-order phase transition. The corresponding
energy release can be estimated from~(\ref{jumpvc2}) as $ {\delta
F(v_{c2})}\approx \frac{47 a_0^2}{8b_0} t^{4/3}\simeq 5.9 \,
t^{4/3}{( T_c\Delta C_p)}$, where $\Delta C_p=0.76\cdot 10^7\,{\rm
erg/(cm^3 K)}$ is the specific heat jump at $T_c$~\cite{GS}.

\section{Rotating superfluids. Pulsars}

The novel phase with the condensate of excitations may also exist
in rotating systems.  Excitations can be generated because of the
rotation. Presence of friction with an external wall or difference
between velocities of the superfluid and the normal fluid are not
necessarily required to produce excitations. Now we should use the
angular momentum conservation instead of the momentum
conservation. The structure of the order parameter is more
complicated than the plane wave~\cite{Voskresensky:1993uw}. With
these modifications, the results, which we obtained above for the
motion with the constant $\vec{v}$, continue to hold. The value of
the critical angular velocity $\Omega_{c1} \sim v_{c1}/R$
is very low for systems of a large size $R$.

In the inner crust and in a part of the core of a neutron star,
protons and neutrons are paired in the 1S$_0$ state owing to
attractive $pp$ and $nn$ interactions. In denser regions of the
star interior the 1S$_0$ pairing disappears but neutrons might be
paired in the 3P$_2$ state~\cite{Sauls89}. The charged superfluid
should co-rotate with the normal matter without forming vortices,
this results in the appearance of a tiny magnetic field
$\vec{h}=2m_p\vec{\Omega}/e_p$ (London effect) in the whole volume
of the superfluid, $m_p$ ($e_p$) is the proton mass
(charge)~\cite{Sauls89}. This tiny field, being $\lsim 10^{-2}$G
for the most rapidly rotating pulsars, has no influence on the
relevant physical quantities and can be neglected.

With the typical neutron star radius, $R\sim 10$\,km, and for
$\Delta\sim$MeV typical for the $1S_0$ $nn$ pairing, we estimate
$\Omega_{c1}\sim 10^{-14}$~Hz. For $\Omega>\Omega_{c1}$ the
neutron star contains arrays of neutron vortices with regions of
the superfluidity among them, and the star as a whole rotates as a
rigid body. The vortices would cover the whole space, only if
$\Omega$ reaches unrealistically large value $\Omega_{c2}^{\rm
vort}\sim 10^{20}$~Hz.  The most rapidly rotating pulsar PSR
J1748-2446ad has the angular velocity
4500~Hz~\cite{Manchester:2004bp}. The value of the critical
angular velocity for the formation of the condensate of
excitations in the neutron star matter  is $\Omega_c\sim
\Omega_c^{\rm L}\simeq \Delta/(p_{\rm F}R)\sim 10^2$~Hz for the
pairing gap $\Delta \sim$ MeV and $p_{\rm F}\sim 300$\,MeV$/c$ at
the nucleon density $n\sim n_0$, where $n_0$ is the density of the
atomic nucleus, and $c$  is the speed of light. The
superfluidity continues to coexist with the condensate of
excitations and the array of vortices until the rotation frequency
$\Omega$ reaches the value $\Omega_{c2}>\Omega_c^{\rm L}$, at
which both the condensate of excitations and the superfluidity
disappear completely. From eq.~(\ref{tildet}) with the BCS
parameters we estimate $\Omega_{c2}\sim v_{c2}/R\lsim 10^4$ Hz.
Thus, in the detected rapidly rotating pulsars the
condensate of excitations might coexist with superfluidity.

\section{Conclusion}

In this letter we studied a possibility of the condensation of
excitations with $k\neq 0$, when a superfluid flows with a
velocity larger than the Landau critical velocity, $v>v_c^{\rm
L}$. We included an interaction between the ``mother'' condensate
of the superfluid and the condensate of excitations and considered
the superfluid at zero and finite temperatures. We assumed that
the superfluid and normal components move with equal velocities.
In practice it might be achieved for $v>v_{c1}$, where $v_{c1}$ is
the critical velocity of the vortex formation. We found that the
condensate of excitations appears in a second-order phase transition
at $v=v_c^{\rm L}$ and the condensate amplitude grows linearly
with the increasing velocity. Simultaneously the mother condensate
decreases and vanishes at $v=v_{c2}$, then the superfluidity is
destroyed in a first-order phase transition with an energy
release. For $v_{c}^{\rm L}<v<v_{c2}$ the resulting flow velocity
is $v_{\rm fin}\le v_c^{\rm L}$, whereby the equality is realized
for $T=0$.
We argued that for the fermion superfluids the
condensate of bosonic excitations might be stable against the appearance of fermionic excitations from the Cooper-pair breaking.
Finally, we considered condensation of excitations
in rotating superfluid systems, such as pulsars. Our estimates
show that superfluidity might coexist with the condensate of
excitations in the rapidly rotating pulsars.

\acknowledgments

The work was supported by Grants No. VEGA~1/0457/12 and No. APVV-0050-11,
by ``NewCompStar'', COST Action MP1304, and by Polatom ESF network.


\end{document}